\documentclass[page-classic,linenumbers]{epl2} 

\usepackage{amsmath, amssymb}
\usepackage[english]{babel}
\usepackage{graphicx, subfigure, float}
\usepackage{epstopdf}
\usepackage{url}

\newcommand{\Eq}[1]{Eq. (\ref{#1})}
\newcommand{\Fig}[1]{Fig. \ref{#1}}

\newcommand{\quant}[2]{#1 $\mathrm{#2}$}

\DeclareMathOperator{\tr}{tr}

\title{Optimal Estimation of Recurrence Structures from Time Series}
\shorttitle{Optimal Estimation of Recurrence Structures}

\author{
    Peter beim Graben\inst{1} \and
    Kristin K. Sellers\inst{2,3} \and
    Flavio Fr\"{o}hlich\inst{2,3} \and
    Axel Hutt\inst{4}
    }
\shortauthor{P. beim Graben \etal}

\institute{
  \inst{1} Bernstein Center for Computational Neuroscience Berlin, Germany \\
  \inst{2} Department of Psychiatry, University of North Carolina at Chapel Hill, USA \\
  \inst{3} Neurobiology Curriculum, University of North Carolina at Chapel Hill, USA \\
  \inst{4} German Meteorological Service, Offenbach am Main, Germany
}


\abstract{
Recurrent temporal dynamics is a phenomenon observed frequently in high-dimensional complex systems and its detection is a challenging task. Recurrence quantification analysis utilizing recurrence plots may extract such dynamics, however it still encounters an unsolved pertinent problem: the optimal selection of distance thresholds for estimating the recurrence structure of dynamical systems. The present work proposes a stochastic Markov model for the recurrent dynamics that allows to derive analytically a criterion for the optimal distance threshold. The goodness of fit is assessed by a utility function which assumes a local maximum for that threshold reflecting the optimal estimate of the system's recurrence structure. We validate our approach by means of the nonlinear Lorenz system and its linearized stochastic surrogates. The final application to neurophysiological time series obtained from anesthetized animals illustrates the method and reveals novel dynamic features of the underlying system. As a conclusion, we propose the number of optimal recurrence domains as a statistic for classifying an animals' state of consciousness.
}

\begin{document}

\maketitle

\section{Introduction}
\label{sec:intro}

Complex behavior is ubiquitous in nature, humanities and engineering. Often, complex dynamical systems are \emph{recurrent} in the sense that certain regions of their available state space are frequently visited in the course of time \cite{Poincare1890, MarwanEA07}. This important property facilitates forecasting, modeling and control of dynamical systems. To visualize recurrent behavior, Eckmann et al. \cite{EckmannOliffsonRuelle87} suggested the recurrence plot (RP) technique that inspired the increasing research field of recurrence quantification analyses (RQA). RP and RQA found several applications in the physical sciences \cite{Casdagli97, SouzaVianaLopes08}, medicine \cite{JavorkaTrunkvalterova08, LiSleighEA07}, social sciences \cite{KaragianniKyrtsou11, OyaAiharaHirata14} and engineering \cite{HuniczEA13, CarrionMirallesLara14} (for further surveys, see \cite{MarwanEA07, Marwan11}).

For a $d$-dimensional time series $\{ \vect{x}_t \}, \vect{x}_t \in \mathbb{R}^d, t=1, \ldots, N$, with $N$ the number of time steps, the RP is a graphical representation of the recurrence matrix
\begin{equation}\label{eq:rp}
    R_{ij} = \Theta(\varepsilon -  || \vect{x}_j - \vect{x}_i ||) .
\end{equation}
The parameter $\varepsilon$ is the distance threshold of two sampling points $\vect{x}_i, \vect{x}_j$ and $\Theta$ denotes the Heaviside step function. Proper selection of the threshold $\varepsilon$ is of crucial importance to gain instructive RPs that reveal the underlying recurrence structure of the dynamics~\cite{MarwanEA07, Marwan11}.

To solve the threshold selection problem, several heuristics have been suggested \cite{MarwanEA07, Marwan11, GrabenHutt13, GrabenHutt15}. However, a proper optimization method should take the system's specific recurrence structure into account by explicitly modeling the alternating sequence of \emph{recurrence domains} and \emph{transients}. One viable way to do so, models the percolation transition toward the giant cluster in a recurrence network \cite{DongesEA12}. Another possibility is the detection of recurrence domains though \emph{recurrence grammars} \cite{GrabenHutt13, GrabenHutt15}. A recurrence grammar comprises rules $i \to j$  for the RP elements when $i > j$ and $R_{ij} = 1$, applied recursively to the sequence of time indices $s_t = t$. This map yields a symbolic dynamics $s'$ which is a coarse-graining of the system's trajectory $\vect{x}_t$; it transforms the recurrence structure into an alternating sequence of recurrence domains and transients. In this Letter, we present a model for the system's recurrence structure in terms of a Markov chain obtained from its symbolic dynamics $s'$. Under the assumption that the optimal Markov chain describes transitions between several metastable states and a transient regime \cite{GrabenHutt15}, we formulate three plausible criteria for the shape of the stochastic transition matrix leading to a novel utility function. Its maximization over a range of distance thresholds $\varepsilon$ yields then an optimal estimate of the system's recurrence structure.

\section{Method}
\label{sec:method}

\begin{figure}[H]
\centering
\includegraphics[scale=0.3]{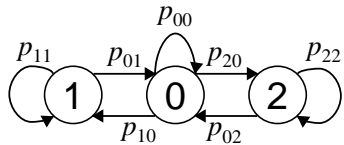}
\caption{\label{fig:markov} Optimal Markov chain for recurrence grammar of the Lorenz attractor \cite{Lorenz63}. State 0 denotes the transient regime acting as a ``hub''. The metastable states 1 and 2, corresponding to the attractor's ``wings'', have considerably large dwells reflected by large self-transition probabilities $p_{ii}$.}
\end{figure}

For easy comparison with our previously presented maximum entropy criterion \cite{GrabenHutt13, TosicEA16}, we consider the Lorenz system \cite{Lorenz63} as a well-known example for a system with a distinguished recurrence structure. Intuitively, the system exhibits two recurrence domains, namely the attractor's ``wings'' centered around its unstable foci and separated by several non-recurrent regimes, henceforth referred as to ``transients''. Therefore, in an optimal encoding the symbolic sequence $s'$ obtained from a recurrence grammar would look like
\begin{equation}\label{eq:ops}
    s' = 0 0 \dots 0 1 1 \dots 1 0 0 \dots 0 2 2 \dots 2 0 0 \dots 1 1 1 \dots 1 0 0 \dots 0 2 2 \dots 2 0 0 \dots
\end{equation}
after applying the rewriting grammar \cite{GrabenHutt13, GrabenHutt15, TosicEA16}. I.e. we expect a number of 0's for some transient regime, followed by a relatively large number of 1's for the first wing, followed by a smaller number of 0's, again, characterizing the transition to the second wing, which is hence reflected by a relatively large number of 2's, and so on. This symbolic dynamics is essentially a Markov chain with a rather simple transition graph depicted in \Fig{fig:markov}.

\vspace{3cm}
\vspace{1cm}

Figure \ref{fig:markov} points out three important properties of the expected optimal encoding: (\emph{i}) The two recurrence domains (states 1 and 2) and the transient state 0 are modeled as metastable states with relatively large dwells, i.e. they exhibit mainly self-transitions. Hence the transition probabilities $p_{00}, p_{11}$, and $p_{22}$ will be large compared to other transitions. (\emph{ii}) The transient state 0 acts as a ``hub''. Without further knowledge about the system's dynamics, it is reasonable to assume that the transition probabilities $p_{10}, p_{20}$ from the hub and $p_{01}, p_{02}$ into the hub are uniformly distributed according to a maximum entropy principle \cite{GrabenHutt13}. (\emph{iii}) There a no direct transitions between metastable recurrence domains 1 and 2. This model can be easily generalized to a higher number of recurrent states.

For a chosen threshold $\varepsilon$, the recurrence grammar yields a segmentation of the time series into $n(\varepsilon)$ symbols. Therefore, the appropriate model is an $n$-state Markov chain with number of recurrence domains (NRD) $n - 1$. The transition matrix for an optimal recurrence grammar partition would be for $n > 1$
\begin{equation}\label{eq:tranmat}
    \vect{P} = \begin{bmatrix}
            1 - (n - 1)q    &    r       &    r       &     \cdots       &     r \\
            q                   & 1 - r    &    0       &     \cdots       &     0 \\
            q                   & 0    &    1 - r       &     \cdots       &     0 \\
            \hdotsfor{5} \\
            q                   & 0    &    0       &     \cdots       &     1 - r  \\
    \end{bmatrix}
\end{equation}
For $n = 1$ the systems exhibits only the transient state 0 and we set $\vect{P} = 1$. The matrix $\vect{P}$ reflects the three optimization criteria formulated above as follows: (\emph{i}) The trace $\tr \vect{P} = 1 + (n - 1)(1 - q - r)$ dominates all other transition probabilities. (\emph{ii}) The transition probabilities from the ``hub'' $q = p_{i0}$ and into the ``hub'' $r = p_{0j}$ for $i, j > 0$ are uniformly distributed. (\emph{iii}) As $\vect{P}$ is a stochastic matrix the column sums are $1 - (n - 1)q + (n - 1) q = 1$ and $r + (1 - r) = 1$, the latter term excluding direct transitions between metastable states.

The three optimization criteria eventually lead to the desired utility function: (\emph{i}) Maximization of the recurrent self-transitions is achieved by maximizing the trace of the transition matrix $\tr \vect{P}$. (\emph{ii}) The maximum entropy principle is satisfied by renormalization of transition probabilities of the first row and of the first column of $\vect{P}$ after neglecting $p_{00}$, i.e., $p_{0j}' = p_{0j} / \sum_{j=1}^{n-1} p_{0j}$ for the first row and $p_{i0}' = p_{i0} / \sum_{i=1}^{n-1} p_{i0}$ for the first column. Then
\begin{eqnarray}\label{eq:1entropy}
  h_r &=& -\frac{1}{\log(n - 1)} \sum_{j=1}^{n-1} p_{0j}' \log p_{0j}'  \nonumber \\
  h_c &=& -\frac{1}{\log(n - 1)} \sum_{i=1}^{n-1} p_{i0}' \log p_{i0}' \:.
\end{eqnarray}
For the optimal Markov matrix \Eq{eq:tranmat} $h_r = h_c = 1$ (for $n = 1$ we trivially define $h_r = h_c = 0$, as there are no transitions). (\emph{iii}) Simultaneously maximizing the trace and the entropies of the first row and column of $\vect{P}$ also suppresses transitions between any two recurrence domains due to the normalization condition for stochastic matrices with $\sum_{i=0}^{n-1} p_{ij} = 1$. Finally, in the limit $q, r \to 0$, the optimal transition matrix in \Eq{eq:tranmat} turns into the unit matrix $\vect{I}$ with $\tr \vect{I} = n$. Consequently, the utility function
\begin{equation}\label{eq:util}
    u(\varepsilon) = \frac{1}{n + 2} \Big[ \tr \vect{P}(\varepsilon) + h_r(\varepsilon) + h_c(\varepsilon) \Big]
\end{equation}
 with $0\le u(\varepsilon)\le 1$ allows us to define the optimization criterion
\begin{equation}\label{eq:optcod}
    \varepsilon^* = \arg \max_\varepsilon u(\varepsilon) \:.
\end{equation}
The parameter $\varepsilon^*$ is the optimal threshold for which the recurrence structure detected from the time series resembles most the Markov chain model. In addition, one may call $u(\varepsilon^*)$ the degree of resemblance of the time series' recurrence structure with the Markov chain model. The number of recurrence domains (NRD), $n(\varepsilon^*) - 1$, characterizes the ``complexity'' of the metastable transition model.

For illustration, consider two interesting limiting cases. As long as the threshold $\varepsilon$ remains smaller than the smallest distance between two sampling points, all points are isolated and hence encoded as transients 0, such that $n = 1$ and $u(\varepsilon \to 0) = 1/3$. Moreover, for thresholds $\varepsilon$ larger than the largest distance between two samples, all points are merged into one single recurrence domain 1. However, the transient state 0 must always be present for reasons of consistency but is not realized in this case, such that $\vect{P}$ becomes the 2-dimensional unit matrix. This leads to $u(\varepsilon \to \infty) =1/2$.

Optimal recurrence grammar encoding permits the detection of temporal segments in neurophysiological time series that best fit to recurrence domains \cite{TosicEA16}. The NRD serves as a measure of complexity in further evaluation. Recent studies \cite{Schartner_etal15,966601} have provided evidence that the complexity of electroencephalographic data measured under general anaesthesia reflects the level of consciousness of subjects. In order to evaluate the optimal recurrence grammar encoding in this context, we investigate spatially distributed Local Field Potentials (LFP) measured in ferret visual cortex under anaesthesia (cf.~\cite{Sellers_etal13} for all details). Briefly, we performed electrophysiological recordings in primary visual cortex in one adolescent female ferret in a dark room while the animal was head-fixed, first while awake and later when anesthetized. Electrophysiology was conducted with acute invasive insertion of 32-channel probes (\quant{50}{\mu m} with contact spacing along the $z$-axis and the reference electrode located on the same shank \quant{0.5}{mm} above the top recording site). Unfiltered signals were amplified, then digitized at a rate of \quant{20}{kHz} and finally down sampled to a rate of \quant{100}{Hz}. For anesthetized recordings, general anesthesia was maintained with isoflurane ($0.5\%$, $0.75\%$ or $1.0\%$) and continuous infusion of xylazine. At least 15 minutes elapsed after changing anesthetic concentration prior to starting a new recording, exceeding the duration required in our setup for the LFP to stabilize at the new anesthetic concentration. All three levels of anesthesia used in this study corresponded to lack of behavioral response. We have extracted a large set of trials, each lasting \quant{10}{s}, for each condition. According to previous studies, anaesthesia strongly affects neural activity in the $\alpha$-frequency band from \quant{8}{Hz} to \quant{12}{Hz} \cite{Sellers_etal13,Purdon_etal12}. Consequently, we bandpass-filtered the data in the $\alpha$-band and extracted instantaneous power by a Gabor filter at each electrode. This yields a $32$-dimensional signal that serves as the input signal of our recurrence analysis. In addition, to test for linearity we have phase-randomized all time series in order to generate surrogate data which we analyzed identically to the original data.

\section{Results}
\label{sec:results}

\begin{figure}[H]
\centering
\subfigure[]{\includegraphics[scale=0.4]{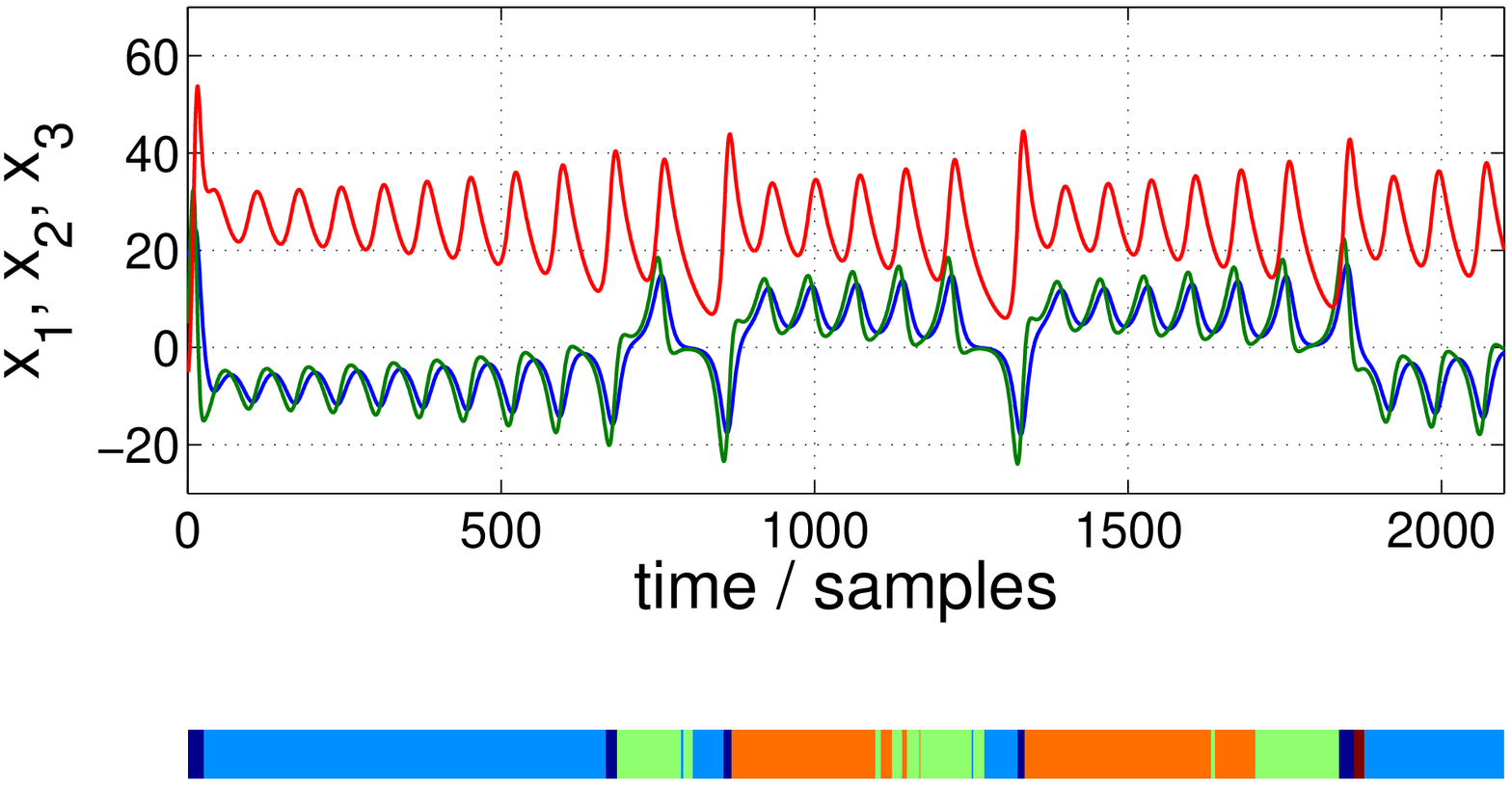}}
\subfigure[]{\includegraphics[scale=0.4]{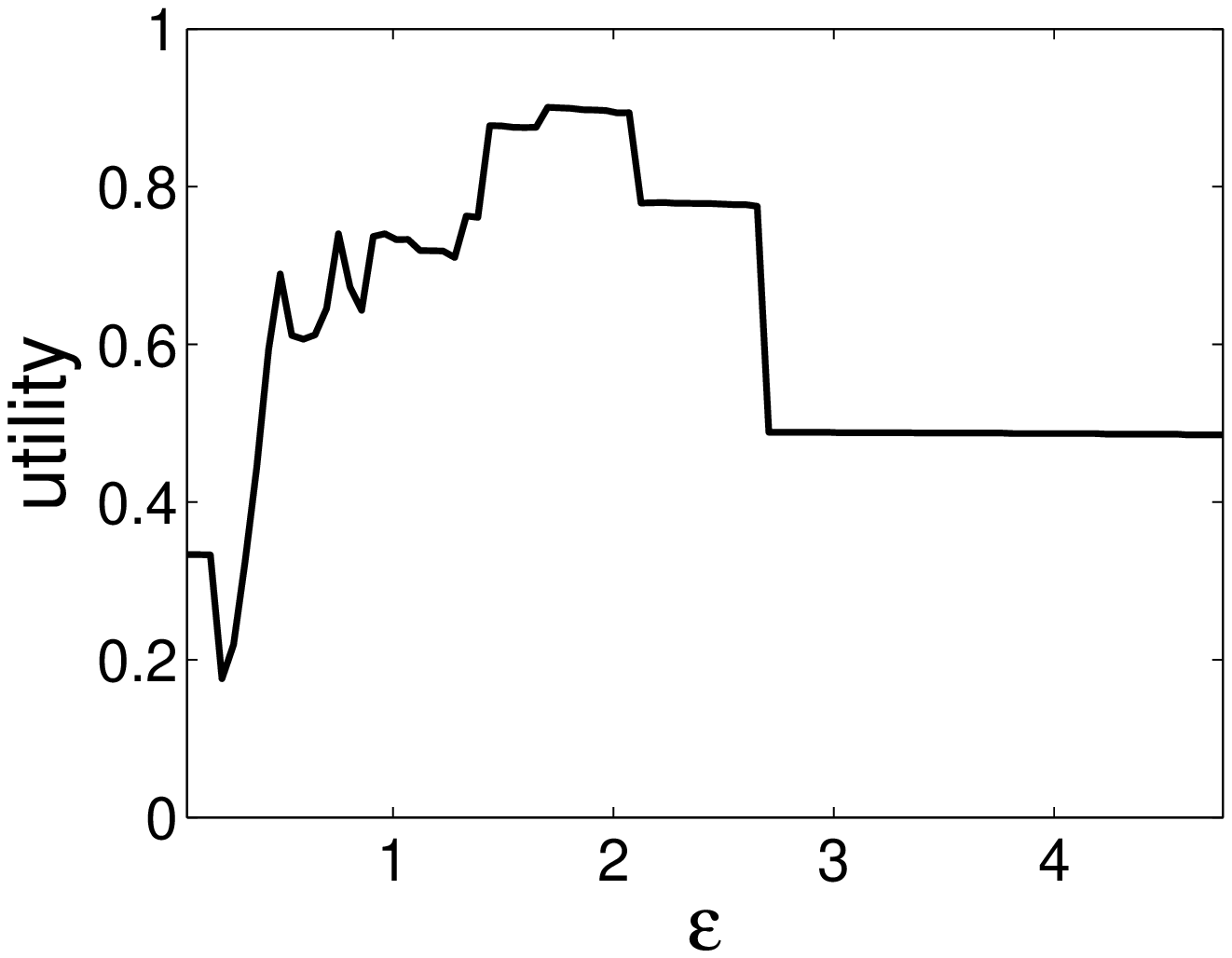}}
\subfigure[]{\includegraphics[scale=0.4]{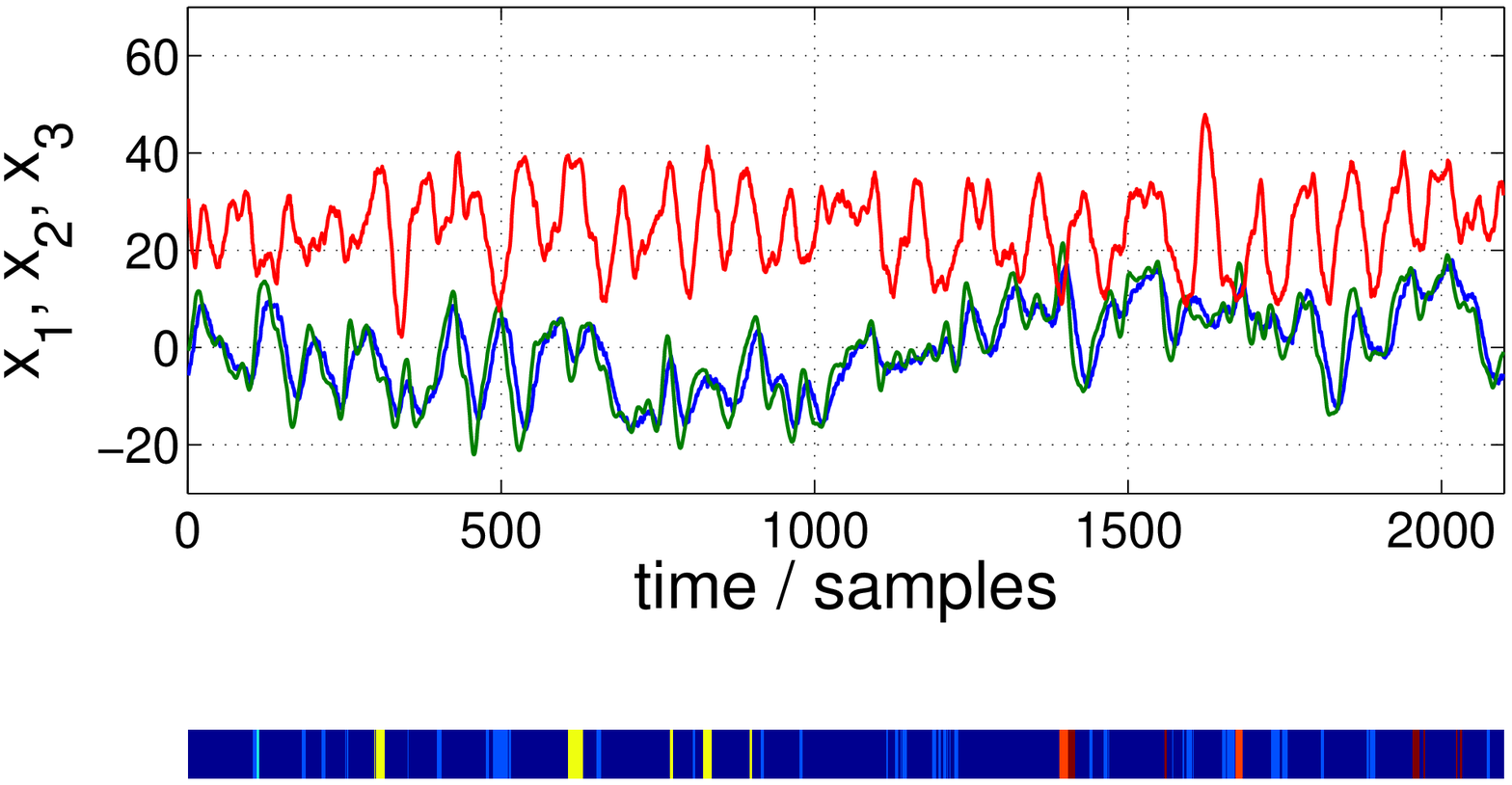}}
\subfigure[]{\includegraphics[scale=0.4]{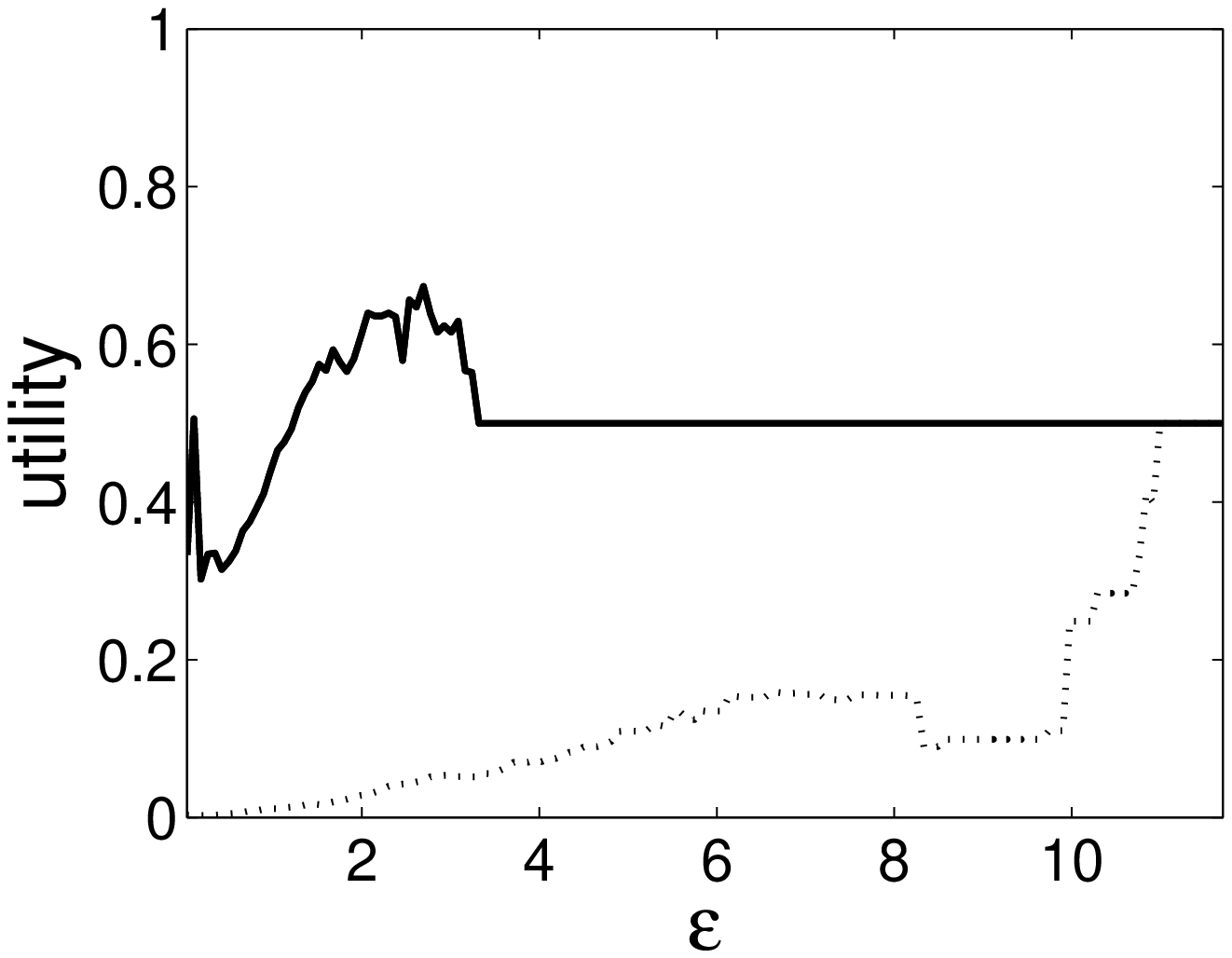}}
\caption{\label{fig:lorenz} (Color online) (a, b) Optimal recurrence grammar partition of the Lorenz attractor \cite{Lorenz63}. (a) Time series $\vect{x}_t$ (upper panel) and optimal encoding $s'$ (color bar beneath). (b) Markov utility function $u(\varepsilon)$ [\Eq{eq:util}]. (c, d) Optimal recurrence grammar partition of Lorenz surrogates. (c) Phase randomized surrogates $\vect{x}'_t$ (upper panel) and corresponding  optimal encoding $s'$ (color bar beneath). (d) Markov utility function $u(\varepsilon)$ [\Eq{eq:util}] for phase randomized (solid) and shuffled surrogates (dotted).}
\end{figure}

In order to validate the new optimization procedure, we  consider again the standard Lorenz system with the same numerical settings as in~\cite{GrabenHutt13}. For each $\varepsilon$, we compute the recurrence plot, its symbolic sequence and the corresponding transition probabilities between symbols (see external Supplement,\footnote{
    We provide an external Supplement at \url{https://www.researchgate.net/publication/301525222_GrabenEA_EPLsupp}
} Sec. 2). Figure \ref{fig:lorenz} presents the results. The upper panel of \Fig{fig:lorenz}(a) displays the time series of the model coordinates $x_1, x_2$ (blue and green) and $x_3$ (red) of the Lorenz trajectory $\vect{x}_t$. The Markov utility function \Eq{eq:util} in \Fig{fig:lorenz}(b) assumes its maximum  at $\varepsilon^* = 1.7$ (compare with $\varepsilon^* = 1.9$ in \cite{GrabenHutt13}). Using this threshold for the optimal recurrence grammar encoding of the trajectory $\vect{x}_t$ yields the color-coded symbolic dynamics $s'$ depicted in the bottom panel of \Fig{fig:lorenz}(a). We observe a 5-state Markov chain with $4$ recurrence domains. The segmentation into the two Lorenz wings as recurrence domains is clearly visible validating the optimal estimation of the system's recurrence structure.

Nonlinear dynamical systems exhibit nontrivial temporal recurrence structures. To further evaluating the proposed method we compare the recurrence structure of the Lorenz system with recurrence structures in linear systems. To this end, we create two kinds of linear stochastic surrogates through shuffling and phase-randomization. Both preserve the statistical distribution of the data but destroy any nonlinear dependencies of the original time series, although only the latter retains the signal's linear autocorrelation structure \cite{PrichardTheiler94}.

Figure~\ref{fig:lorenz}(c,d) shows the Lorenz surrogate data. Obviously, the maximum of the utility function for the phase randomized surrogates (\Fig{fig:lorenz}(d: solid)) is substantially suppressed compared to that of the original time series. The optimal recurrence grammar for these surrogates yields a Markov chain with $6$ recurrence domains (as compared to $4$ in the original data), as seen in \Fig{fig:lorenz}(c). These correspond to smooth peaks of the time series that are spuriously detected as saddle nodes \cite{GrabenHutt15}. For comparison, \Fig{fig:lorenz}(d) also gives the utility function for time-shuffled surrogates (dotted line in panel d) mimicking white noise (compare with external Supplement, Sec. 1). The function $u(\varepsilon)$ approaches $u(\varepsilon^*) = 1/2$ as its maximum that corresponds to a single recurrence domain. This result illustrates that a system without any essential recurrence structure is described by a regular sequence of only one recurrence domain, as in the case of white noise (external Supplement, Sec. 1).

\begin{figure}[H]
\centering
\subfigure[]{\includegraphics[scale=0.35]{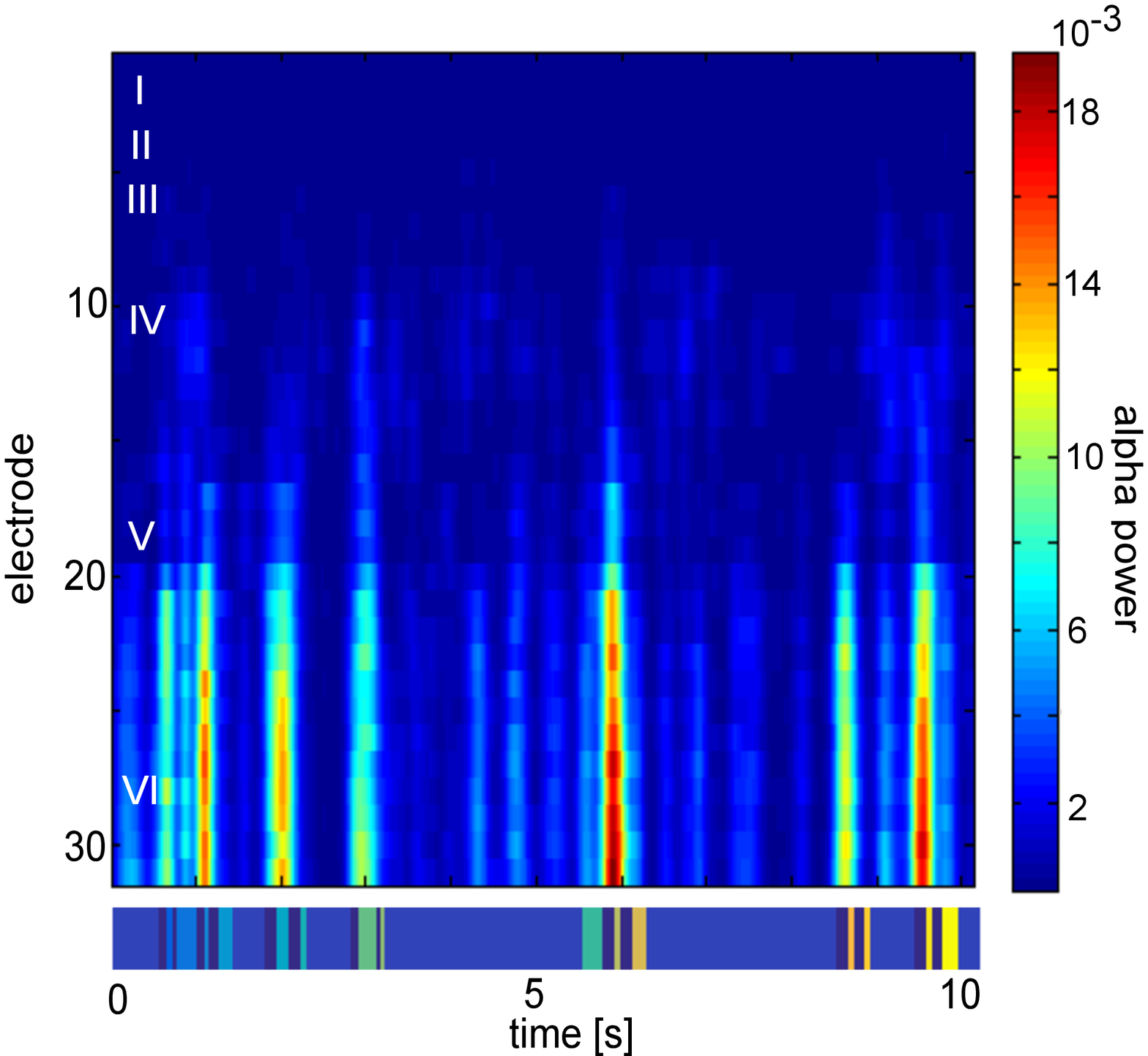}}
\subfigure[]{\includegraphics[scale=0.35]{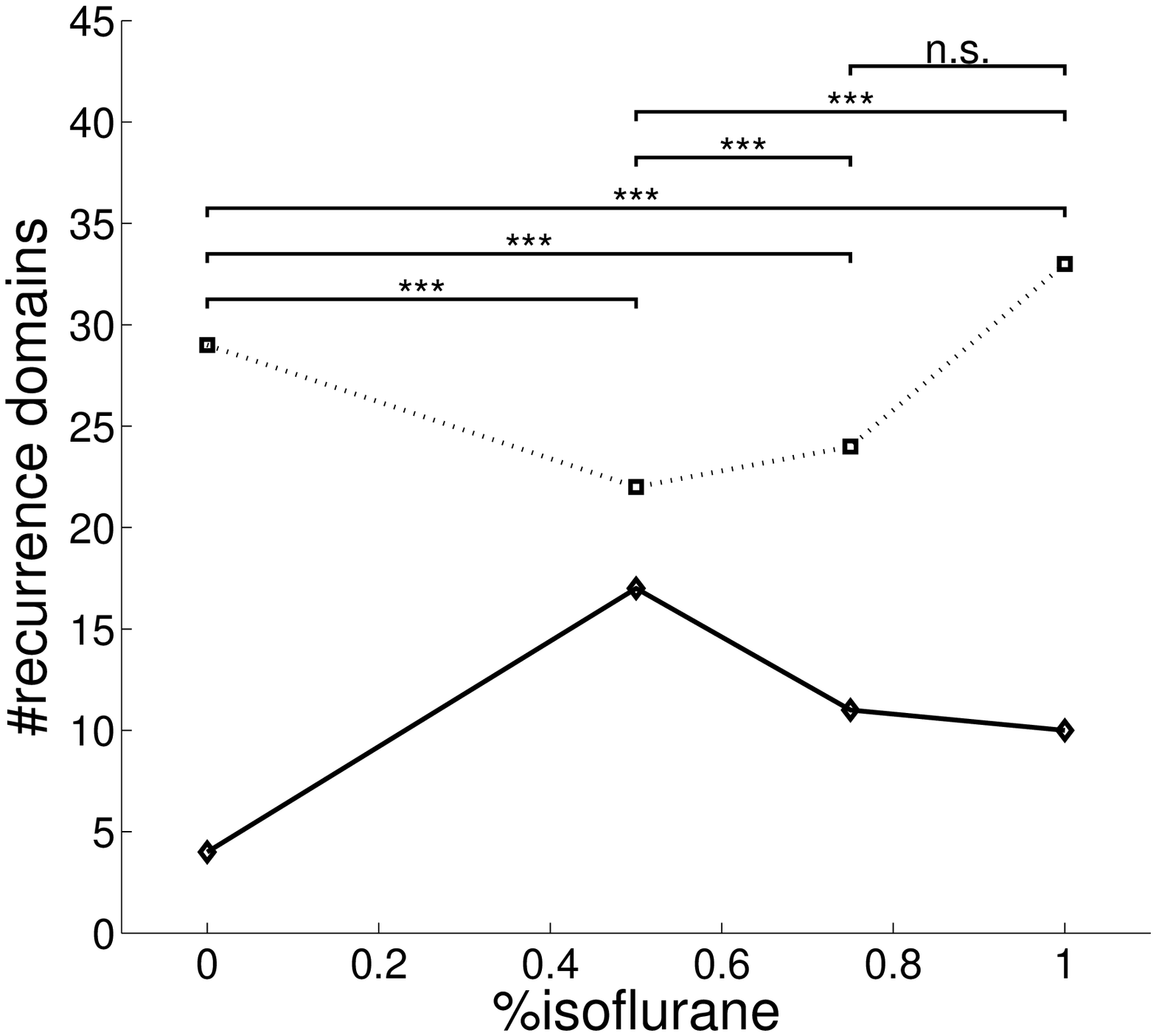}}
\caption{\label{fig:c05} Neural activity in the $\alpha$-frequency band for various isoflurane concentrations.
(a) Time-resolved spatial distribution of power in a single trial under isoflurane concentration of $0.5\%$ (top) where electrode 1 to 6 denote the top cortical layers $I-III$, electrode 7-14 the medium layer $IV$ and $15-32$ the bottom layers $V-VI$. The corresponding color-coded optimal symbolic sequence $s^\prime$ shows good accordance to the distributed activity (bottom). The trial is selected to have the medium number of recurrence domains for $0.5\%$ concentration, cf. panel on the right.
(b) Median number of recurrence domains subject to the isoflurane concentration for the original data (solid line) and
phase-randomized surrogates (dotted line). The data sets comprise number of trials of $132$ ($0\%$), $193$ ($0.5\%$), $180$ ($0.75\%$) and $176$ ($1.0\%$), together with pairwise statistical comparisons using a Kolmogorov-Smirnov test.}
\end{figure}

Next, Figure \ref{fig:c05}(a) shows the instantaneous $\alpha$-power of the original single trial LFP from an animal anesthetized with $0.5\%$ isoflurane concentration in $32$ channels distributed over space together with the optimal recurrence sequence. We observe a sequence of spatially distributed activity states in the $32$-dimensional data (top panel) and the gained optimal sequence of symbolic states (bottom panel) that shows good accordance to the experimental data (similar representations in external Supplement, Sec. 4). Hence the optimal recurrence grammar encoding identifies well the present spatiotemporal sequences. We determined the distribution of NRD over all trials for each level of anesthesia (illustrated in the external Supplement, Secs. 2 and 4). These distributions are bimodal and thus not normal. Hence, \Fig{fig:c05}(b) displays medians for both the original and surrogate time series. Firstly, we observe a strong dependence of NRD from the level of anesthesia which is significant for almost every pairwise comparison using a Kolmogorov-Smirnov test ($p < 10^{-6}$) between all concentrations in the original data set, except the difference between $0.75\%$ and $1.0\%$ ($p > 0.05$). Moreover, results from original and phase-randomized data are group-wise significantly different (Kruskall-Wallis test, $p < 0.05$) demonstrating that the optimal recurrence grammar encoding clearly distinguishes nonlinear from linear dynamics. The awake state ($0\%$) exhibits much less recurrence domains than the sedation states with  isoflurane concentrations larger than $0.5\%$. This reflects less different recurrent states in the brain, repetitions of fewer states and perhaps a higher degree of transient dynamics in the awake state compared to the anesthetized states. After the increase of NRD from $0\%$ to $0.5\%$ reflecting less regular signals, the significant decrease of NRD from $0.5\%$ to $0.75\%$ and $1.0\%$ indicates a progress towards higher regularity, i.e., ordering under deeper anaesthesia. The change of regularity may result from changes of intrinsic noise levels in neural populations. In fact, increased concentrations of anaesthetic agents diminish the activity input from sub-cortical to cortical structures leading to more regular neural activity~\cite{Hudetz+Alkire_in_MashourBook11}. This final decrease of regularity is in line with the well-established classification of EEG of anaesthesia: the deeper anaesthesia, the more regular or synchronous is the observed EEG. Our result in the $\alpha$-frequency range shows a maximum in NRD indicating a new non-homogeneous relation between anaesthetic concentration and regularity.

\section{Discussion}
\label{sec:disc}

The present work proposes a novel optimal estimate of recurrence structures that allows to characterize nonlinear temporal structures in multivariate time series by the number of recurrence domains. In contrast to previously suggested heuristics for the optimization of recurrence thresholds \cite{MarwanEA07, Marwan11, GrabenHutt13, GrabenHutt15}, our approach explicitly models the system's recurrence structure as transitions in a Markov chain between metastable states and transient regimes. We obtain a small number of recurrence domains with relatively large dwells that can be related to transitive subnetworks in a recurrence network close to the percolation transition \cite{DongesEA12}. The proposed method also shows similarities with other methods for finding a system's \emph{generating partition} though partitioning homoclinic connections \cite{GrassbergerKantz85}, as metastable states could be segmented along heteroclinic connections. We leave this interesting issue for future research.


\acknowledgments

This research has been supported by the European Union's Seventh Framework Programme (FP7/2007-2013) ERC grant agreement No. 257253 awarded to AH and by a Heisenberg fellowship (GR 3711/1-2) of the German Research Foundation (DFG) awarded to  PbG. The research reported in this publication was partially supported by the National Institute of Mental Health of the National Institutes of Health under Award Number R01MH101547. The content is solely the responsibility of the authors and does not  necessarily represent the official views of the National Institutes of Health.



\end{document}